\documentclass[aps, prl,10pt,notitlepage,nofootinbib,superscriptaddress,showkeys,twocolumn]{revtex4-1}
\usepackage{amstext,amsmath,amssymb,amsfonts,bbm, amsthm}
\usepackage[latin1]{inputenc}
\usepackage{fancyhdr}
\usepackage{graphicx}
\usepackage{hyperref}
\usepackage{tocvsec2}

\usepackage{color}

\def\beq{\begin{equation}}
\def\be{\begin{equation}}
\def\ee{\end{equation}}
\def\bes{\begin{eqnarray}}
\def\ees{\end{eqnarray}}

\DeclareMathOperator{\tr}{tr}

\def\f{\frac}

\begin{document}
%%%%%%%%%%%%%%%%%%%%%%%%%%%%%%%%%%%%%%%%%%%%%%%%%%%

\title{\large \bf Universality in $p$-spin glasses with correlated disorder}

\author{{Valentin Bonzom}}\email{vbonzom@perimeterinstitute.ca}
\author{{Razvan Gurau}}\email{rgurau@perimeterinstitute.ca}
\affiliation{Perimeter Institute for Theoretical Physics, 31 Caroline St. N, ON N2L 2Y5, Waterloo, Canada}
\author{{Matteo Smerlak}}\email{smerlak@aei.mpg.de}
\affiliation{Max-Planck-Institut f\"ur Gravitationsphysik, Am M\"uhlenberg 1, D-14476 Golm, Germany}

\date{\small\today}

%%%%%%%%%%%%%%%%%%%%%%%%%%%%%%%%%%%%%
\begin{abstract}\noindent
We introduce a new method, based on the recently developed \emph{random tensor theory}, to study the $p$-spin glass model with non-Gaussian, correlated disorder. Using a suitable generalization of Gurau's theorem on the universality of the large $N$ limit of the $p$-unitary ensemble of random tensors, we exhibit an infinite family of such non-Gaussian distributions which leads to \emph{same} low temperature phase as the Gaussian distribution. While this result is easy to show (and well known) for uncorrelated disorder, its robustness with respect to strong quenched correlations is surprising. We show in detail how the critical temperature is renormalized by these correlations. We close with a speculation on possible applications of random tensor theory to finite-range spin glass models.

%In strict analogy with the 't Hooft expansion of large random matrices, a suitable ensemble of random \emph{tensors} has recently been found to possess a well defined large $N$ limit, dominated by a simple family of Feynman diagrams. What is more, the large-$N$ correlation functions in this ensemble turn out to be \emph{universally Gaussian}. Here we show that, in the context of (infinite-range) $p$-spin glass models, this property translates into the universality of the spin-glass transition for an infinite class of non-Gaussian \emph{correlated} disorder distributions: all perturbations in this class change the critical temperature but not the structure of the low temperature phase. We also speculate on possible applications of random tensor theory to finite-range spin glass models.
\end{abstract}
%%%%%%%%%%%%%%%%%%%%%%%%%%%%%%%%%%%%%%
\maketitle

\paragraph{Introduction.}

It is well known that the phase diagram of the $p$-spin glass model \cite{Derrida1981,Crisanti1992} does not depend on the details of the disorder distributions, in the following sense: if $J_{i_{1}\cdots i_{p}}$ denotes a set of independent and identically distributed set of $p$-valent coupling between sites $i_{1}\cdots i_{p}$, a non-quadratic potential $V(J_{i_{1}\cdots i_{p}})$ in the coupling distribution
\be
\prod_{i_{1}\cdots i_{p}}dJ_{i_{1}\cdots i_{p}}\,e^{-J_{i_{1}\cdots i_{p}}^{2}/\sigma^{2}+V(J_{i_{1}\cdots i_{p}})}
\ee
is irrelevant in the thermodynamic limit. That a similar result would hold for \emph{correlated} disorder distributions, with terms such as
\be\label{example}
\sum_{\{i_l,j_l\}}J_{i_1 i_2 i_3}J_{i_1 j_2 j_3} J_{j_1 j_2 j_3}J_{j_1 i_2 i_3},
\ee
in the potential, is much less obvious. In fact, to our knowledge, no analytic framework to deal with such correlated, non-Gaussian disorder has been reported so far. Since disorder correlations are to be expected in actual physical systems, understanding their effect is an important problem.

In this Letter we exhibit an infinite class of non-Gaussian terms of the kind \eqref{example} such that $(i)$ the thermodynamic limit $N\rightarrow\infty$ is exactly soluble, and $(ii)$ the spin glass phase has the same structure as with uncorrelated disorder, except for a renormalization of the critical temperature. This provides the first general result on spin glasses with strongly correlated disorder.

Our approach is based on new results in \emph{random tensor theory}. As natural generalizations of random matrices, random tensors have recently been showed  to possess a large $N$ limit \cite{Gurau2011b} dominated by few, well-identified ``melonic'' graphs (the tensor equivalent of 't Hooft's planar graphs in matrix theory \cite{Hooft:1973jz}). Furthermore, the melonic family can actually be resummed exactly, and turns out to exhibit interesting critical and multicritical behavior \cite{Bonzom2011b}. These results have not been applied to spin-glass problems previously, and our hope is to convey that random tensors are potentially as powerful tools for spin glass theory as random matrices \cite{Kosterlitz1976}.

From the perspective of random tensor theory, the quenched couplings of spin glasses with $p$-spin interactions are non-Gaussian rank-$p$ random tensors. The behavior of such tensors in the large $N$ limit has been investigated in \cite{Gurau2011e, Bonzom:2012hw}, with a striking conclusion: in a suitable ensemble with \emph{$p$-unitary} symmetry (more details in the text), this limit is \emph{universally Gaussian}. This means that, in this ensemble, in the large $N$ limit the sole effect of the self-interactions of large tensors is to dress the propagator. Here, we show how this result can be generalized to include interactions between tensors and spin variables, and thus obtain the aforementioned universality result.

This Letter is organized as follows. We first recall the Hamiltonian for $p$-spin models and insist on the need for a correlated disorder. Then, we recall the relevant properties of large random tensors in the $p$-unitary ensemble. This enables us to show how non-Gaussian, correlated, quenched variables can be integrated exactly in the large $N$ (thermodynamic) limit, yielding our universality theorem. We conclude with a few words on the possible relevance of tensor techniques for short-range $p$-spin glasses.

\paragraph{$p$-spin glass models.}
We consider a $p$-spin Hamiltonian \cite{Derrida1981,Crisanti1992}
\be\label{ham}	
H_{J}(S)=-\sum_{1\leq i_{1}\cdots i_{p}\leq N}J_{i_{1}\cdots i_{p}}S_{i_{1}}\cdots S_{i_{p}}+c.c.
\ee
where $J_{i_1\dotsb i_p}$ is a complex\footnote{The use of complex rather than real tensors is motivated by purely technical convenience and does not change the physics in any way.} tensor describing the couplings and
$S=(S_{i})_{1\leq i\leq N}$ is a set of real spins\footnote{It is also possible to include Potts or vector spins coupled according to some fixed multi-linear map.} with lattice index $i$, weighted by a (normalized) probability measure $d\Omega(S)$ such that
\be\label{spincondition}
\int d\Omega(S)\,\sum_{i=1}^{N}S_{i}^{2}=\mathcal{O}(N).
\ee
This includes in particular Ising \cite{Derrida1981} and spherical \cite{Crisanti1992} spins.

When the couplings are Gaussianly distributed, such $p$-spin glass models are well-known to exhibit replica symmetry breaking in the low temperature phase \cite{Crisanti1992,Gross1984,Gardner1985} and have a dynamical transition at a higher temperature where a large number of metastable states (growing exponentially with $N$) dominates the free energy landscape \cite{Crisanti1995,Crisanti2005}; their relevance is conjectured to extend to structural glasses \cite{Kirkpatrick1987}. These results extend easily to the case of independent and identically distributed (i.i.d.) couplings: all terms of higher order than $2$ in $J_{i_{1}\dots i_{p}}$ and $\overline{J}_{i_{1}\dots i_{p}}$ in the measure on $(J,\overline{J})$ are irrelevant in the thermodynamic limit $N\rightarrow\infty$.

% assume that the couplings $J_{i_{1}\cdots i_{p}}$ are distributed in the $p$-unitary invariant ensemble defined above, which consists of \emph{non-Gaussian} and \emph{non-independent} (i.e. \emph{correlated}) couplings.

In this Letter we aim to study a family of \emph{correlated} non-Gaussian measures on the disorder. Physically, randomness of the couplings comes from randomness of the positions of the spins, and in general we should not expect the couplings between different sets of $p$ spins to be independent (for instance due to the geometric relations between the positions of the spins). One should therefore perturb the Gaussian distribution (with covariance $\sigma^2$) on the couplings with a polynomial $V(J,\overline{J})$. The quenched free energy is given by
\be
[F(J,\overline{J})] = \frac{\int dJ d\overline{J} e^{-N^{p-1}\left(J\cdot\overline{J}/\sigma^{2}+V(J,\overline{J})\right)} F(J,\overline{J})}{\int dJd\overline{J}\, e^{-N^{p-1}\left(J\cdot\overline{J}/\sigma^{2}+V(J,\overline{J})\right)}},
\ee
with
\be
-\beta F(J,\overline{J}) = \ln \int d\Omega(S) e^{-\beta H_{J}(S)},
\ee
and $J\cdot \overline{J}$ is shorthand for $\sum_{j_{1}\dots j_{p}}J_{j_{1}\dots j_{p}}\overline{J}_{j_{1}\dots j_{p}}$.

The evaluation of $[F]$ for a generic potential $V$ is of course a completely open problem. However, the theory of large random tensors provides an exact calculation for an infinite family of potentials satisfying a particular kind of invariance.

\paragraph{Large random tensors.}
We now review the relevant properties of large random tensors discovered in \cite{Gurau2011b,Bonzom2011b,Gurau2011e}. The first obvious observation is that, unlike symmetric/hermitian matrices, tensors cannot be diagonalized. Hence, a key concept in random matrix theory, the eigenvalue distribution, does not carry over to the higher-rank case. It turns out however that this fact does not preclude the development of random tensor theory, which in fact relies on the identification of an \emph{ensemble} with suitable symmetry properties.

\begin{figure}
 \includegraphics[scale=0.5]{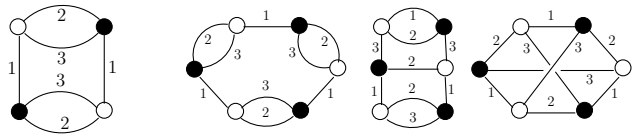}
\caption{\label{fig:T4invariant} Some $p$-bubbles at $p=3$. Up to color re-labeling there is a single bubble with 4 vertices (on the left), whose invariant is $ \sum_{\{i_l,j_l\}}J_{i_1 i_2 i_3} \overline{J}_{i_1 j_2 j_3} J_{j_1 j_2 j_3} \overline{J}_{j_1 i_2 i_3}$. But there exist different bubbles with 6 vertices (the three on the right).}
\end{figure}

One such ensemble of tensors---indeed the only one identified so far---is the \emph{$p$-unitary ensemble}, defined as follows. Consider a rank-$p$ tensor in $N$ complex dimensions $J$, with components $J_{i_1\dots i_{p}}$ in a fixed basis, and for each set of $p$ unitary matrices $U^{(1)}$ to $U^{(p)}$, define
\be
J'_{i_1\dots i_{p}}=\sum_{j_{1}\dots j_{p}}U^{(1)}_{i_{1}j_{1}}\cdots\, U^{(p)}_{i_{p}j_{p}}\,J_{j_1\dots j_{p}}.
\ee
Then we say that a function $V(J,\overline{J})$ of $J$ and its complex conjugate $\overline{J}$ is a \emph{$p$-unitary} invariant if\footnote{The corresponding symmetry group is known as the external tensor product of $p$ copies of $U(N)$.}
\be
V(J',\overline{J'})=V(J,\overline{J}).
\ee
The set of $p$-unitary invariants is conveniently parametrized by \emph{$p$-bubbles} $B$, that is $p$-valent bipartite connected graphs with edges colored by numbers between $1$ and $p$, such that each ``color'' is incident exactly once to each vertex, see Fig. \ref{fig:T4invariant}. A bubble represents an invariant denoted $\tr_B(J, \overline{J})$, by associating a tensor $J$ to each ``white'' vertex of $B$ and a conjugate $\overline{J}$ to each ``black'' vertex, and contracting their $k$-th indices along the edges colored by $k$. By the fundamental theorem of classical invariants of $U(N)$ (see for instance \cite{U(N)invariants}), a general $p$-unitary invariant can be expanded as
\be\label{defpotential}
V(J,\overline{J})=\sum_{B}t_{B}\tr_{B}(J,\overline{J}),
\ee
where $t_{B}$ are coupling constants.

For a given invariant potential $V(J,\overline{J})$, we define the average of $f(J,\overline{J})$ over $J$ by
\be\label{tensoraverage}
\Big[f(J,\overline{J})\Big]=\f{\int dJd\overline{J}\, e^{-N^{p-1}\left(J\cdot\overline{J}/\sigma^{2}+V(J,\overline{J})\right)} f(J,\overline{J})}{\int dJd\overline{J}\, e^{-N^{p-1}\left(J\cdot\overline{J}/\sigma^{2}+V(J,\overline{J})\right)}}.
\ee
The Feynman diagrammatic expansion of these quantities involves $(p+1)$-colored bipartite graphs, made of $p$-bubbles connected together via extra lines with color ``0'' incident on each vertex and corresponding to the propagator $\sigma^{2}$ in \eqref{tensoraverage}.

\begin{figure}
 \includegraphics[scale=0.4]{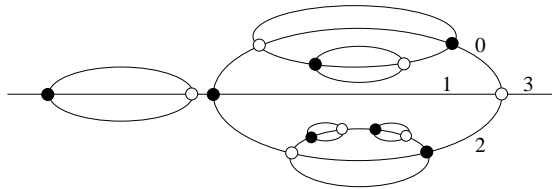}
 \caption{\label{fig:melon} A patch of a melonic graph, recursively built by inserting on any line a pair of vertices, a black and a white, connected together via $p$ lines (here $p=3$). }
\end{figure}

The following results concerning the large $N$ limit of \eqref{tensoraverage} have been proved:

\begin{itemize}
\item
The Feynman expansion is dominated in the large $N$ limit by a simple class of graphs, called \emph{melonic graphs}, which generalize 't Hooft's planar graphs \cite{Gurau2011b}. Intuitively, a $(p+1)$-colored graph is melonic if it can be built by recursive insertions on any line of two vertices connected together by $p$ lines, as in Fig. \ref{fig:melon}.
%\medskip
\item
The large $N$ limit is \emph{Gaussian}, in the sense that up to subleading corrections in $1/N$,
\be\label{factorization}
\Big[\tr_{B}(J,\overline{J})\Big]= NG_2^{|B|/2},
%\Big[\tr_{B}(J,\overline{J})\Big]=\Big[\tr_{B}(J,\overline{J})\Big]_{0},
\ee
where $|B|$ is the number of vertices of the bubble $B$ and $G_2 = [ J\cdot \overline{J}]/N$ is the \emph{dressed} propagator depending on the potential $V$ \cite{Gurau2011e}.
%where $[\,\cdot\,]_{0}$ denotes a Gaussian average with a \emph{dressed} propagator depending on the potential $V$ \cite{Gurau2011e}.
%\medskip
\item
The following Schwinger-Dyson equation holds in the $N\rightarrow\infty$ limit \cite{Bonzom:2012hw}
\be\label{SD}
\frac{[J\cdot \overline{J}]}{\sigma^2\,N}+\sum_{B}t_{B}\f{\vert B\vert}{2}\f{\big[\tr_{B}(J,\overline{J})\big]}{N}=1,
\ee
%where $\vert B\vert$ is the number of vertices of $B$ \cite{Bonzom:2012hw}.
\end{itemize}
The first result implies that all non-melonic bubbles $B$ in the potential drop out in the large $N$ limit, and therefore we can restrict the sum in \eqref{defpotential} to melonic bubbles (hence hereafter $B$ will always denote a melonic bubble). In Fig. \ref{fig:T4invariant}, all bubbles are melonic except the non-planar one on the right.

The second result has been coined the \emph{universality} property of the $p$-unitary ensemble of random tensors, and can be seen as a non-trivial generalization of the central limit theorem. Its origin is that there is only one way to dress a melonic bubble $B$ with propagators in a melonic way, which happens to correspond to Gaussian contractions. This feature is specific to tensors and does not hold for random matrices. In a way, this Letter can be read as the physics counterpart of this surprising mathematical result. We refer the reader to the review \cite{Gurau2012} and to the original papers for more details on random tensor theory.

%\paragraph{The KTJ method for the $p$-spin spherical model.}
%Let us come back to spin glass theory. In this section, we consider the Crisanti-Sommers $p$-spin spherical model, with Hamiltonian
%\be
%H_{J}[S]=\sum_{1\leq i_{1}\cdots i_{p}\leq N}J_{i_{1}\cdots i_{p}}S_{i_{1}}\cdots S_{i_{p}}+c.c.,
%\ee
%where each $1\leq i\leq N$ is a site index, $S_{i}$ are \emph{complex} spherical spins, viz. $\sum_{i=1}^{N}\vert S_{i}\vert^{2}=N$, and assume for the moment a Gaussian distribution for complex quenched couplings $J=(J_{i_{1}\cdots i_{p}})$, i.e. $V=0$ in the language of the previous section. The use of complex variables rather than real ones is a purely technical assumption, which does not change the physics.\footnote{It should be possible to use real spins directly; however, since the theory of random tensors has been developed in the complex, we prefer to stick to this choice.} We now report our attempt to mimick the KTJ ansatz-free approach in the $p$-spin case.

\paragraph{Universality in the couplings.} Let us now come back to spin glasses. Following the standard recipe to compute quenched quantities \cite{Mezard1986}, we consider the averaged replicated partition function
\be
\big[Z^{n}\big]=\int\prod_{a=1}^{n}d\Omega(S^{a})\, e^{-\beta H_{\textrm{eff}}(\{S^{a}\})},
\ee
where $a$ is the replica index and the effective Hamiltonian is defined by \be
e^{-\beta H_{\textrm{eff}}(\{S^{a}\})}=\Big[e^{-\beta \sum_{a=1}^{n}H_{J}(S^{a})}\Big].
\ee
In diagrammatic language, $H_{\textrm{eff}}$ is given by the sum over all connected $(p+1)$-colored bipartite graphs (henceforth ``graph'') with spins $S^{a}_{i}$ on the external legs. Denoting $k$ the order of the effective coupling between replicas $a_{1}\cdots a_{k}$, this can be pictured as

%\begin{widetext}
\be\label{dessin}
-\beta H_{\textrm{eff}}(\{S^{a}\})=\sum_{k}\beta^{k}\sum_{a_1\dotsc a_k} \begin{array}{c}\includegraphics[scale=0.65]{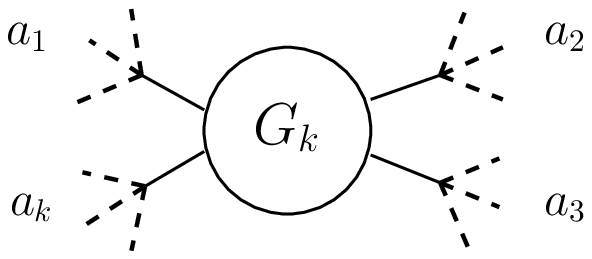}\end{array}
\ee%\end{widetext}
Here the solid line is the $J$-propagator (tensor lines with color $0$), and the $p$ dashed line emerging from each external leg represents the external spin variables $S^{a_{k}}_{i_{l}}$. The blob $G_{k}$ is the large-$N$ tensor connected $k$-point function, i.e. the sum over all connected melonic graphs with $k$ external (solid) legs. For each graph contributing to the blob amplitude, the site indices $i_{l}$ of the spins are contracted along ``broken faces'', i.e. connected paths with alternating color $1\leq c\leq p$ and $0$ from one external dashed leg to another through the graph.

Let us now show that $k=2$ terms dominates in the large $N$ limit. Observe that powers of $N$ in $H_{\textrm{eff}}(\{S^{a}\})$ have three sources: the tensor propagators, the bubble interactions $\tr_{B}(J,\overline{J})$, and the sums over site indices $i$. The first two contributions are those of a melonic graph with $k$ cut lines of color $0$ whose scaling has been was found in the appendix of \cite{Gurau2011e} to be $p-(p-1)k-\rho$, where $\rho$ is a positive number independent of $k$. As for the spin contribution, from \eqref{spincondition} we see that it gives at most a factor of $N$ per broken face, and there are at most $pk/2$ of them. This gives for the scaling degree $\omega(k)$ in $N$ of the order-$k$ term of \eqref{dessin}
\be
\omega(k)\leq p-(\f{p}{2}-1)k.
\ee
We conclude that, indeed, only $k=2$ terms are relevant in the large $N$ limit. Thus, at leading order \eqref{dessin} reduces to
\be
-\beta H_{\textrm{eff}}(\{S^{a}\})= \beta^2 \sum_{a,b} \begin{array}{c}\includegraphics[scale=0.65]{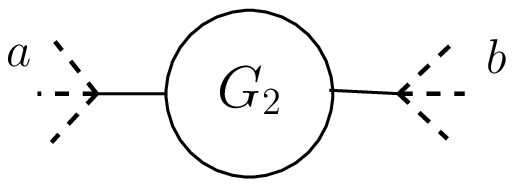}\end{array}
\ee

To complete our evaluation of the effective Hamiltonian, we must compute the $2$-point function $(G_2)_{i_1\dotsb i_p;j_1\dotsb j_p} = [J_{i_1\dotsb i_p} \overline{J}_{j_1\dotsb j_p}]$ of the tensor. Its scaling with $N$ is $N^{-(p-1)}$. Its tensorial structure is $\prod_{l=1}^p \delta_{i_l,j_l}$ which identifies by pairs the lattice sites between the replicas $a$ and $b$. Finally its amplitude, simply denoted  $G_2$, is found by inserting the universality property \eqref{factorization} into the Schwinger-Dyson equation \eqref{SD}, yielding
%\be
%v(x)= \frac{x}{\sigma^2} + \sum_{m}\left(\sum_{B\in\mathcal{B}_{m}}t_{B}\right)x^{m},
%\ee
%in which $\mathcal{B}_{m}$ denotes the set of melonic bubbles with $2m$ vertices, this gives
%\be\label{twopoint}
%G_{2}\,v'(G_{2})=1.
%\ee
\be\label{twopoint}
\frac{G_2}{\sigma^2} + \sum_{m\geq 2} \biggl(\sum_{B\in\mathcal{B}_{m}}t_{B}\biggr) m\, G_2^m =1,
\ee
in which $\mathcal{B}_{m}$ denotes the set of melonic bubbles with $2m$ vertices. The leading-order connected $2$-point function is the solution of this polynomial\footnote{Or analytic, if $V$ has infinitely many bubble terms.} equation, and depends on the whole set of coupling constants $t_{B}$. For example, for a potential with a single $4$-vertex bubble (see Fig. \ref{fig:T4invariant}) with coupling constant $t$, equation \eqref{twopoint} becomes
\be
2\sigma^2 tG_{2}(t)^{2}+G_2(t)=\sigma^2,
\ee
hence, picking the solution with $G_{2}(0)=\sigma^{2}$,
\be
G_{2}(t)=\f{\sqrt{1+8\sigma^4 t} -1}{4\sigma^2 t}.
\ee
This is a smoothly decreasing function of $t\geq -1/8\sigma^{4}$.

Summarizing, we have proved that
\be
-\beta H_{\textrm{eff}}(\{S^{a}\})=\frac{\beta^{2}G_{2}}{N^{p-1}}\sum_{a,b}\sum_{i_{1}\cdots i_{p}}\,\prod_{l=1}^{p}S^{a}_{i_{l}} S^{b}_{i_{l}},
\ee
which is the usual $p$-spin replica Hamiltonian \cite{Gross1984,Castellani2005}, except for the variance $\sigma^{2}$ which is replaced by $G_{2}$ (which as we saw can be computed exactly for a given tensor quenched potential $V$). This is the content of our \emph{universality theorem}, the main result of this Letter. It shows that the higher order terms in the quenched distribution change the critical temperature, but not the structure of the low temperature phase.

\paragraph{Conclusion and outlook.} We have introduced large random tensors as a new tool for spin glass theory. Using the peculiar scaling behavior of tensors in the $p$-unitary ensemble, we have identified an infinite universality class of infinite-range $p$-spin glasses with non-Gaussian correlated quenched distributions. To our knowledge, this is the first universality theorem of spin glass theory with this level of generality.

%Beyond this universality result, random tensor theory may be relevant to gain further understanding of $p$-spin glasses. Kosterlitz, Thouless and Jones used in \cite{Kosterlitz1976} the random matrix technology to solve the infinite-range spherical $2$-spin model \emph{without} replicas. The new status of random tensor theory have enabled us to apply the same method to the spherical $p$-spin model. Although it does give sensible results in the high-temperature phase, it fails to capture the replica symmetric breaking in the low temperature phase. It appears that the KTJ method applies to $p=2$ thanks to an accidental $O(N)$ rotation invariance in replica space which is lost for $p\geq 3$. The results of this ongoing investigation will be reported elsewhere.

%Beyond this universality result, random tensor theory may be relevant to gain further understanding of the replica formalism itself. For instance, we have tentatively applied the Kosterlitz-Thouless-Jones ansatz-free approach to spherical $p$-spins glasses with $p \geq 3$: although it does give a definite result, this result turns out to be inconsistent with replica symmetry breaking. We expect that the origin of this inconsistency in analytical terms; the results of this ongoing investigation will be reported elsewhere.

We close with a more prospective remark. Just like their Sherrington-Kirkpatrick relatives, the $p$-spin interactions in \eqref{ham} have infinite range, and for this reason $p$-spin glass models are judged (at least partially) unphysical. We expect however that random tensor techniques should be applicable to finite-range models too. Indeed, from the random tensor perspective, a finite-range spin glass model is one for which the $J$-propagator is non-trivial, and in particular depends on the tensor indices of $J$. A typical example of interest here would be
\be
\sigma^2_{i_1\dotsb i_p;j_1\dotsb j_p} = \frac{\prod_{k=1}^{p}\delta_{i_{k}j_{k}}}{\sum_{1\leq l< k\leq p} (i_l-j_k)^2+1},
\ee
which goes to zero when the lattice sites are far away. Such tensor models have already been considered in the context of quantum gravity \cite{Rivasseau2011b}, where they have been called \emph{tensor field theories} (TFT). The key difference between TFT and the simple tensor models considered in this Letter is the appearance of a renormalization flow. %Its physical content is expected as follows. When the lattices sites $i_{l}$ and $i_{k}$ are far from each other, the propagator does not vary and is almost constant. This regime is thus identical to the one considered in this Letter and describes clusters where every spin is coupled to every other. Then the renormalization group generates a flow towards a regime where the covariance varies a lot and which may describe nearest-neighbour couplings between the clusters.
The first renormalizable TFT has been identified in \cite{Geloun2011}, and developments are fast in this area. We expect that these new techniques will prove useful in the difficult field of finite-range spin glass theory.

%Performing the Hubbard-Stratonovitch transformation $S^{a}\mapsto (Q_{ab},\lambda_{ab})$ (where $Q_{ab}$ is the replica overlap matrix and $\lambda_{ab}$ is the corresponding Lagrange multiplier) and looking for saddle points in $\lambda_{ab}$ gives the

\paragraph{Acknowledgements.} The authors thank Romain Mari for his careful reading of the manuscript and Francesco Zamponi for useful comments. Research at Perimeter Institute is supported by the Government of Canada through Industry Canada and by the Province of Ontario through the Ministry of Research and Innovation. The research leading to these results has received funding from the European Union Seventh Framework Programme FP7-People-2010-IRSES under grant agreement 269217.

\bibliographystyle{utcaps}

\end{document}